\documentclass[a4paper]{jpconf}
\usepackage{graphicx}
\usepackage{amsmath}
\usepackage{type1cm}
\usepackage{float}

\begin{document}
\title{Constraining viewing geometries of pulsars with single-peaked $\gamma$-ray profiles using a multiwavelength approach}

\author{A~S Seyffert$^1$, C Venter$^1$, T~J Johnson$^{2,3}$ and A~K Harding$^2$}
\address{$^1$~Centre for Space Research, North-West University, Potchefstroom Campus, Private Bag X6001, Potchefstroom 2520, South Africa}
\address{$^2$~Astrophysics Science Division, NASA Goddard Space Flight Center, Greenbelt, MD 20771, USA}
\address{$^3$~Department of Physics, University of Maryland, College Park, MD 20742, USA}
\ead{20126999@nwu.ac.za}

\begin{abstract}
Since the launch of the Large Area Telescope (LAT) on board the \textit{Fermi} spacecraft in June 2008, the number of observed $\gamma$-ray pulsars has increased dramatically. A large number of these are also observed at radio frequencies. Constraints on the viewing geometries of 5 of 6 $\gamma$-ray pulsars exhibiting single-peaked $\gamma$-ray profiles were derived using high-quality radio polarization data~\cite{Weltevrede}. We obtain independent constraints on the viewing geometries of 6 by using a geometric emission code to model the \textit{Fermi}~LAT and radio light curves (LCs). We find fits for the magnetic inclination and observer angles by searching the solution space by eye. Our results are generally consistent with those previously obtained~\cite{Weltevrede}, although we do find small differences in some cases. We will indicate how the $\gamma$-ray and radio pulse shapes as well as their relative phase lags lead to constraints in the solution space. Values for the flux correction factor ($f_\Omega$) corresponding to the fits are also derived (with errors).

\end{abstract}

\section{Introduction}

We study~6 pulsars (J0631+1036, J0659+1414, J0742$-$2822, J1420$-$6048, J1509$-$5850, and J1718$-$3825) detected by \textit{Fermi} LAT, all exhibiting single-peaked pulse profiles within current statistics at energies $> 0.1$~GeV. These pulsars are also detected in the radio band. Both of these properties aid in constraining the possible solution space for the respective pulsar geometries, as derived from predictions of their light curves (LCs).

A study of the pulsars' geometric parameters, the inclination and observer angles $\alpha$ and $\zeta$, has been performed~\cite{Weltevrede} using the radio polarization and LC data. They constrained the solution space for $\alpha$ and $\beta$ of 5 of the individual pulsars, with the impact angle $\beta=\zeta-\alpha$, using fits to these radio data as well as predictions for the value of the half opening angle, $\rho$, of the radio beam derived from the radio pulse width (e.g., \cite{Gil84}).

The aim of this study is to obtain similar constraints on $\alpha$ and $\zeta$ of the same pulsars using an independent, multiwavelength approach. We use a geometric pulsar emission code (e.g., \cite{Venter09}) to model the \textit{Fermi} LAT and radio LCs, and fit the predicted radio and $\gamma$-ray profiles to the data concurrently, thereby significantly constraining $\alpha$ and $\zeta$. This also allows us to test the consistency of the various approaches used to infer the pulsar geometry.

\section{Model}
\label{sec:model}
We use an idealized picture of the pulsar system, wherein the magnetic field has a retarded dipole structure \cite{Deutsch55} and the $\gamma$-ray emission originates in regions of the magnetosphere (referred to as `gaps') where the local charge density is sufficiently lower than the Goldreich-Julian charge density \cite{GJ69}. These gaps facilitate particle acceleration and radiation. We assume that there are constant-emissivity emission layers embedded within the gaps in the pulsar's corotating frame. The location and geometry of these emission layers determine the shape of the $\gamma$-ray LCs, and there exist multiple models for the geometry of the magnetosphere describing different possible gap configurations.

We included two models for the $\gamma$-ray emission regions in this study, namely the Outer Gap (OG) and Two-Pole Caustic (TPC) models. In both the OG and TPC models (the Slot Gap model \cite{Arons83} may serve as physical basis for the latter), emission is produced by accelerated charged particles moving within narrow gaps along the last open magnetic field lines. In the OG model \cite{CHR86}, radiation originates above the null charge surface (where the Goldreich-Julian charge density changes sign) and \textit{interior to} the last open magnetic field lines. The TPC gap starts at the stellar surface and extends \textit{along} the last open field lines up to near the light cylinder\footnote{The slight difference in transverse polar position of the gaps with respect to the magnetic axis (as motivated by physical models) will result in a systematic shift in best-fit ($\alpha$,$\zeta$)-contours between the two models. See Section~\ref{sec:results}.}, where the corotation speed approaches the speed of light~\cite{Dyks03}. The special relativistic effects of aberration and time-of-flight delays, which become important in regions far from the stellar surface (especially near the light cylinder), together with the curvature of the magnetic field lines, cause the radiation to form caustics (accumulated emission in narrow phase bands). These caustics are detected as peaks in the observed $\gamma$-ray LC~\cite{Dyks03,Morini83}.

We used an empirical radio cone model~\cite{Story07}, where the beam diameter, width, and altitude are functions of the pulsar period $P$, its time derivative $\dot{P}$, and the frequency of observation $\nu$, in order to obtain predictions for the radio LCs. This is different from the approach taken by \cite{Weltevrede}, as we have a different prescription for the radio emission altitude and $\rho$, and we do not use the polarization data.

\section{Method}
Using the above geometric models, we generated LCs as a function of $\alpha$ and $\zeta$, keeping the gap widths and extents constant. Due to the large size of the ($\alpha$,$\zeta$)-space at a $1^\circ$ resolution, it is impractical to search blindly for the optimum LC by eye. We therefore started by generating a so-called atlas of LCs with a $10^\circ$ resolution for each pulsar and first identified candidate LC fits. We then performed refined searches in the regions of ($\alpha$,$\zeta$)-space around those candidate fits at a $5^\circ$ and later $1^\circ$ resolution. This approach of generating all the LCs in a candidate region and comparing them directly, allowed us to obtain constraints on $\alpha$ and $\zeta$. We then inferred contours for $\alpha$ and $\zeta$ for each pulsar and for each geometric model. We lastly used these contours to find the corresponding values of $f_\Omega$ (with errors; see Section~\ref{sec:fom}) predicted by the OG and TPC models.


\subsection{Obtaining the contours}
\label{sec:ObCon}
\begin{figure}[t]
      \begin{minipage}[l]{20.5pc}
	\includegraphics[width=20.5pc]{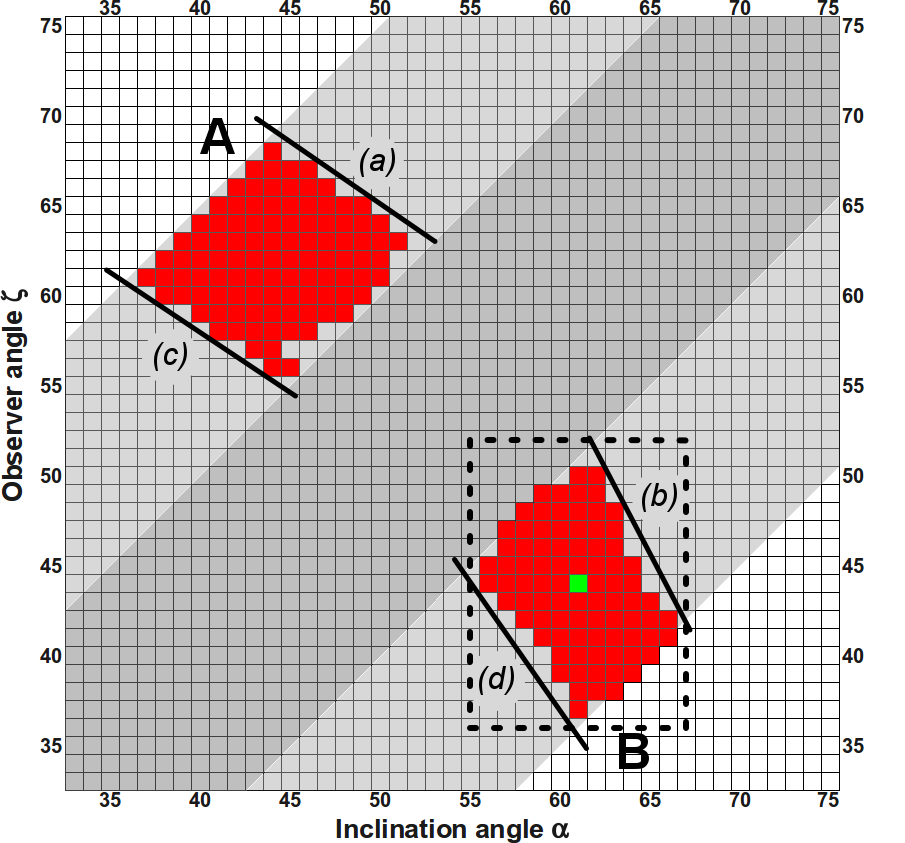}
      \end{minipage}\hspace{1pc}%
      \begin{minipage}{16pc}
	\includegraphics[width=16pc]{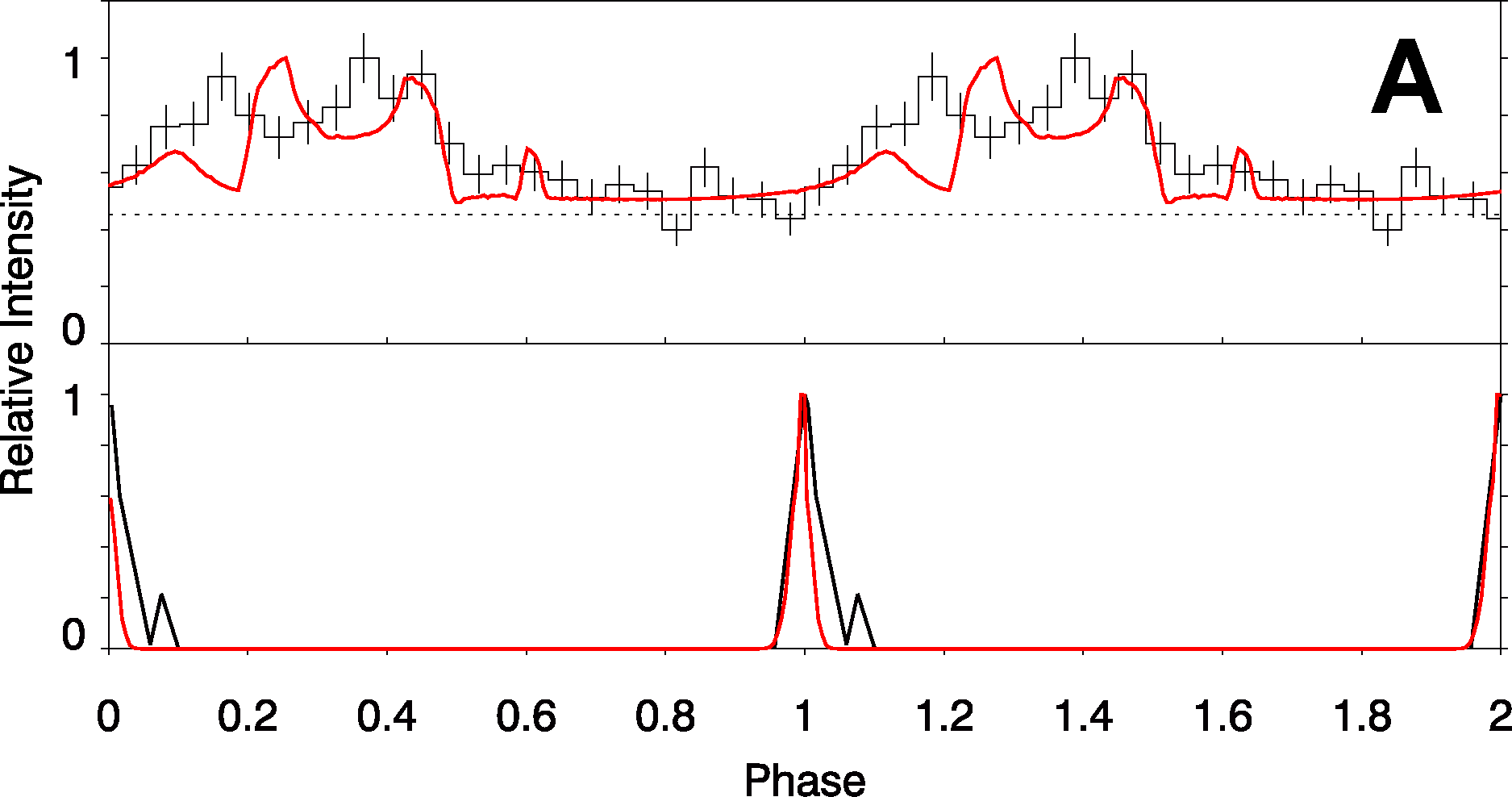}
	\vspace{-0.5pc}
	\newline
	\includegraphics[width=16pc]{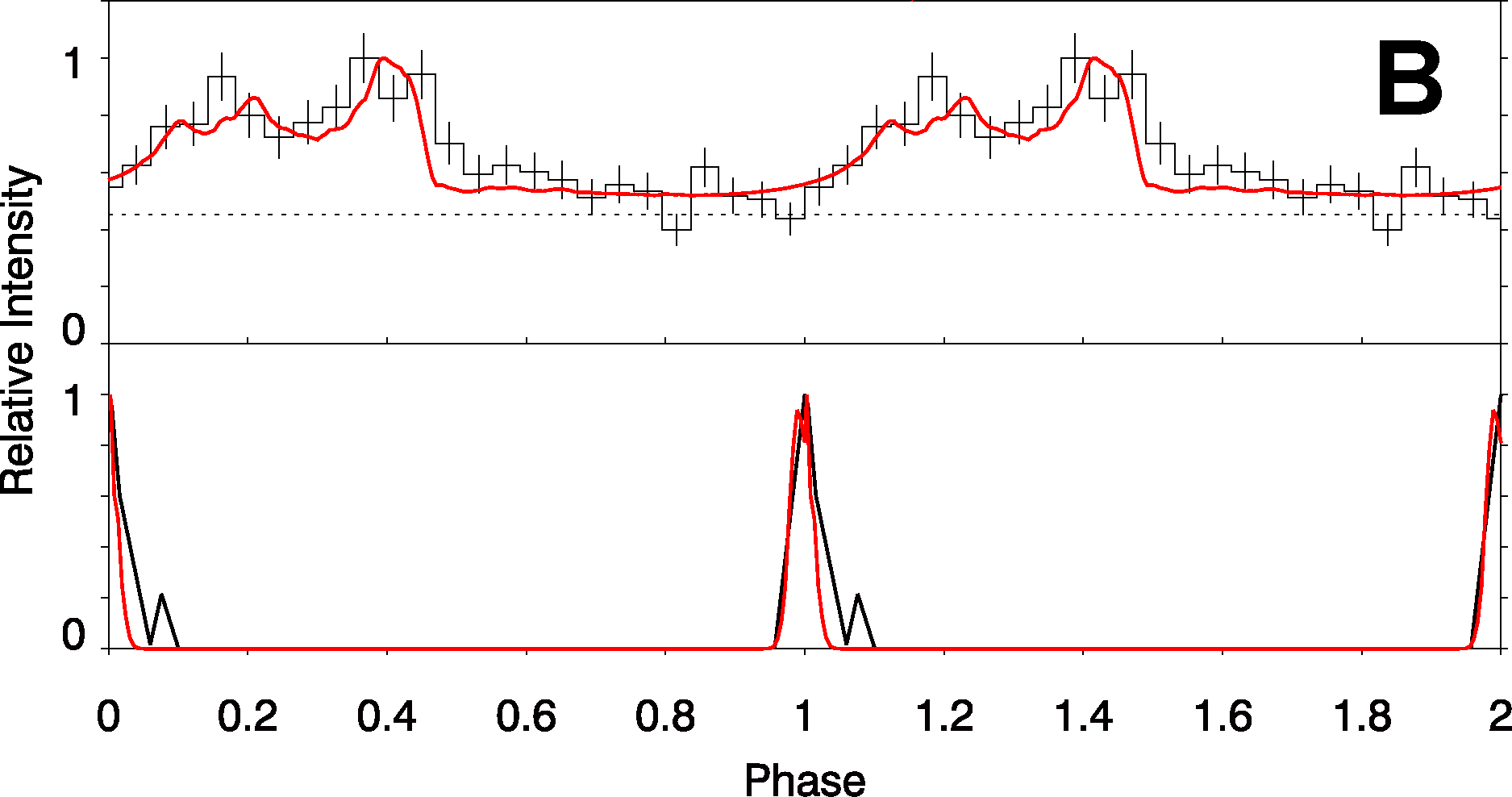}
      \end{minipage}
      \caption{\label{fig:boundConts}\textit{Left:} The ($\alpha$,$\zeta$)-contours (regions \textbf{A} and \textbf{B}) for radio and $\gamma$-ray LC fits of PSR J1509$-$5850 for the TPC model (\textit{red}). The dark grey and light grey regions correspond to double-peaked and single-peaked radio LCs respectively. The solid lines indicate approximately the locations of the boundaries to the contours due to: (\textit{a}) the relative phase lag between the radio and $\gamma$-ray peaks being too large, (\textit{b}) the bridge emission's relative intensity being too low, and (\textit{c,d}) the $\gamma$-ray peak being too narrow. The dashed line (bounded area) indicates the resulting errors on $\alpha$ and $\zeta$, with the \textit{green} block indicating our best fit. \textit{Right:} Representative LC fits corresponding to the contours \textbf{A} and \textbf{B}.}
    \end{figure}

To obtain an ($\alpha$,$\zeta$)-contour delineating the region of plausible LC fits, we start at the candidate solution ($5^\circ$ resolution) and first move away from it in steps of $1^\circ$ in $\alpha$. At each step we then step through $\zeta$ until the upper and lower bounds of the contour are located for the fixed $\alpha$. This procedure is repeated until no further acceptable LCs are found. As an example, the red regions \textbf{A} and \textbf{B} in the left panel of Figure~\ref{fig:boundConts} indicate the resulting contours obtained in this manner for PSR J1509$-$5850 for the TPC case.

Oftentimes we find two disjoint solution contours, with one of the two sometimes producing a better LC fit. The example shown here is one of these cases, as can be seen from the representative LCs shown in Figure~\ref{fig:boundConts}. Contour \textbf{B} yields a considerably better fit than contour~\textbf{A}, and thus we can ignore contour \textbf{A} in favour of contour \textbf{B} as indicated by the dotted line in Figure~\ref{fig:boundConts}. The values of $\alpha$ and $\zeta$ reported in \Tref{tab:results} are those at the centre of this dotted box, while the errors on these values are conservatively chosen to encompass the dimensions of the box. In this case the resulting values are $\alpha=(61^\circ\pm5^\circ)$ and $\zeta=(44^\circ\pm7^\circ)$.

\subsection{Finding $f_\Omega$}
\label{sec:fom}
It is important to be able to convert the observed energy flux of a pulsar to its all-sky luminosity. The flux correction factor $f_\Omega$ is used for this purpose. It is a highly model-dependent parameter, and allows us to determine what fraction of the all-sky luminosity we are observing from a pulsar if the geometry is known ($\alpha$ and $\zeta$), for a given model. It is also crucial in deriving the $\gamma$-ray efficiency of a pulsar, and is defined as~\cite{Watters09}
\begin{equation}
      f_\Omega(\alpha,\zeta_E)=\frac{ \iint {F_\gamma} \left( \alpha , \zeta , \phi \right) \sin \zeta d\zeta d\phi }{ 2 \int {F_\gamma} \left( \alpha , {\zeta}_E , \phi \right) d \phi },
\end{equation}
with $F_\gamma$ being the photon flux per solid angle (`intensity'), and $\zeta_E$ the Earth line-of-sight. The value of $f_\Omega$ is typically taken to be 1, meaning that the observed energy flux is assumed to be equal to the average energy flux over the entire sky. Due to the OG model predicting emission over a relatively smaller region of phase space, its associated $f_\Omega$ values are typically smaller than those of the TPC model for the same $\alpha$ and $\zeta$. (Note, we assume that $F_\gamma(\zeta,\phi)/F_{\rm \gamma,tot}$ is approximately equal to the ratio of observed energy flux vs. total energy flux.)

\begin{figure}[t]
      \begin{center}
	\includegraphics[width=25pc]{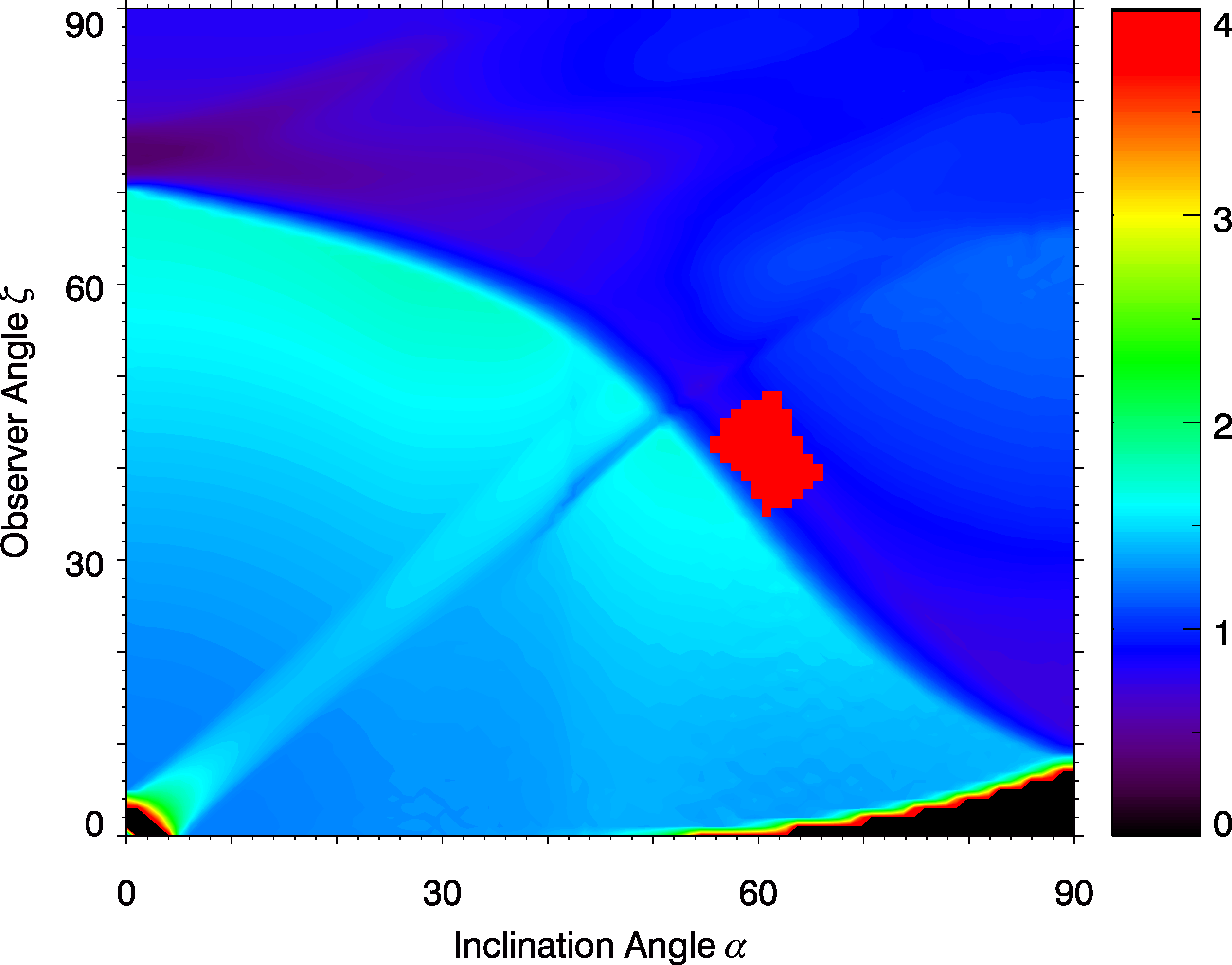}
      \end{center}
      \caption{\label{fig:consOnF}Contour plot of $f_\Omega$ as function of $\alpha$ and $\zeta$ implied by the TPC model for PSR J1509-5850. The \textit{red} area indicates the best-fit $(\alpha,\zeta)$-contour for the radio and $\gamma$-ray LCs.}
    \end{figure}

We can now use the ($\alpha$,$\zeta$)-contours obtained by the method discussed in Section~\ref{sec:ObCon} to constrain the value of $f_\Omega$ by computing $f_\Omega(\alpha,\zeta)$ and overplotting those contours (see Figure~\ref{fig:consOnF}). 

\section{Results}
\label{sec:results}
We infer values for $\alpha$ and $\zeta$ for each pulsar with typical errors of $\sim5^\circ$ (see \Tref{tab:results}). The tabulated values are the average values of $\alpha$, $\zeta$, and $\beta$ implied by the solution contours, while the errors are chosen conservatively so as to include the full (non-rectangular) contour. We note that once a solution is found at a particular $\alpha$ and $\zeta$, it is worthwhile to study the LC at the position $\alpha'=\zeta$ and $\zeta'=\alpha$  (i.e., when the angles are interchanged). Such a complementary solution may provide a good fit in some cases (see the alternative solutions listed in \Tref{tab:results}), due to the symmetry of the model as well the symmetry in pulsar geometry (i.e., $|\beta|$ remaining constant under this transformation).

As mentioned in Section~\ref{sec:model}, we see a shift in best-fit contours, with the TPC model consistently yielding smaller $\alpha$ and $\zeta$ values compared to those obtained using the OG model due to a different gap geometry. This is because similar phaseplots are obtained when a larger $\alpha$ is chosen for the OG model than for the TPC model. Reproduction of the radio pulse shape then necessitates a larger value of $\zeta$ for the OG fit, since the contour is constrained to appear in one of the grey regions (as determined by peak multiplicity; see Figure~\ref{fig:boundConts}). The net effect of this is that the OG contours are found farther from the origin at $\alpha=\zeta=0$ along the grey bands than the corresponding TPC contours.

  \begin{table}[b]
    \caption{\label{tab:results}Values for $\alpha$, $\zeta$, $\beta$, and $f_\Omega$ derived for 6 pulsars for both the OG and TPC models. The $^*$'s indicate alternate best-fit solutions upon interchange of $\alpha$ and $\zeta$.}
    \lineup
    \begin{center}
      \begin{tabular}{llcccc}
	\br
	Pulsar & Model & $\alpha~(^\circ)$ & $\zeta~(^\circ)$ & $\beta~(^\circ)$ & $f_\Omega$\cr
	\mr
	J0631$+$1036 & OG      & $74\pm5\0$ & $67\pm4\0$ & $-\06\pm2\0\m$ & $0.93\pm0.06$\cr
		     & TPC     & $71\pm6\0$ & $66\pm7\0$ & $-\05\pm3\0\m$ & $1.04\pm0.04$\cr
	\mr
	J0659$+$1414 & OG      & $59\pm3\0$ & $48\pm3\0$ & $-12\pm5\0\m$  & $1.16\pm0.53$\cr
		     & TPC     & $50\pm4\0$ & $39\pm4\0$ & $-13\pm6\0\m$  & $1.64\pm0.04$\cr
		     & TPC$^*$ & $38\pm1\0$ & $50\pm4\0$ & $\m11\pm4\0\m$ & $1.63\pm0.05$\cr
	\mr
	J0742$-$2822 & OG      & $86\pm3\0$ & $71\pm5\0$ & $-16\pm6\0\m$  & $0.99\pm0.10$\cr
		     & OG$^*$  & $71\pm6\0$ & $86\pm4\0$ & $\m16\pm6\0\m$ & $0.81\pm0.09$\cr
		     & TPC     & $64\pm8\0$ & $80\pm4\0$ & $\m15\pm6\0\m$ & $0.88\pm0.41$\cr
	\mr
	J1420$-$6048 & OG      & $67\pm5\0$ & $45\pm7\0$ & $-22\pm9\0\m$  & $0.77\pm0.13$\cr
		     & TPC     & $64\pm6\0$ & $43\pm8\0$ & $-21\pm9\0\m$  & $0.90\pm0.10$\cr
		     & TPC$^*$ & $42\pm5\0$ & $63\pm5\0$ & $\m21\pm9\0\m$ & $0.77\pm0.06$\cr
	\mr
	J1509$-$5850 & OG      & $66\pm4\0$ & $50\pm7\0$ & $-18\pm8\0\m$  & $0.77\pm0.11$\cr
		     & TPC     & $61\pm5\0$ & $44\pm7\0$ & $-18\pm8\0\m$  & $0.89\pm0.10$\cr
	\mr
	J1718$-$3825 & OG      & $67\pm6\0$ & $48\pm6\0$ & $-19\pm8\0\m$  & $0.76\pm0.12$\cr
		     & TPC     & $61\pm5\0$ & $43\pm6\0$ & $-19\pm8\0\m$  & $0.86\pm0.07$\cr
		     & TPC$^*$ & $42\pm6\0$ & $62\pm5\0$ & $\m19\pm7\0\m$ & $0.83\pm0.10$\cr
	\br
      \end{tabular}
    \end{center}
  \end{table}

Our results for $f_\Omega$ for each of the pulsars (and models) are given in Table~\ref{tab:results}.

\section{Discussion and Conclusions}
The good constraints on $\alpha$ and $\zeta$, obtained using the method described above, emphasizes the merit of a multiwavelength approach. For example, the light grey regions in Figure~ \ref{fig:boundConts} indicate single-peaked radio profile LC solutions. These would represent typical constraints one would be able to derive for $\alpha$ and $\zeta$ when considering only the radio profile shapes. Similarly, when only considering the (single-peaked) $\gamma$-ray LCs, one would find relatively large ($\alpha$,$\zeta$)-contours~\cite{Watters09}.
The requirement of fitting both the radio and $\gamma$-ray profile shape, as well as their relative phase lag, results in much smaller contours, as shown in Figure~\ref{fig:boundConts}.

Generally, our best-fit ($\alpha$,$\zeta$) compare favourably with those inferred by \cite{Weltevrede} for the first three pulsars (see Table~\ref{tab:results}). However, comparison is hampered by uncertainties in estimating the half opening angle,~$\rho$, which sensitively influences the optimal solutions obtained by \cite{Weltevrede}. Even a small error of $5^\circ$ on~$\rho$ leads to relatively large errors on the allowed $\alpha$, so that our best fits would then be included in their inferred parameter ranges. Direct comparison of our results with those of \cite{Weltevrede} for PSR J1420$-$6048 and PSR J1718$-$3825 is not feasible due to different approaches for estimating the radio pulse width, $W$. We only model the most prominent radio peak, while \cite{Weltevrede} measures $W$ as the width of the total two-peaked profile. Implementing a factor 2 difference in $W$ will change the $\rho$-contours and thus their best-fit solution considerably, improving the agreement between the two approaches. Comparison with \cite{Weltevrede} is not possible for PSR J1509-5850 as the lack of polarization data inhibited inference of a best-fit ($\alpha$,$\beta$) by \cite{Weltevrede} in this case.
 
The next step for this kind of study is to apply a mathematically rigorous method of determining the best-fit LC solutions as a function of many free model parameters to the six pulsars modelled in this paper. An example is the Markov chain Monte Carlo method which has been successfully applied to millisecond pulsar LCs~\cite{Johnson11}.


\newpage
\ack
CV is supported by the South African National Research Foundation. AKH acknowledges support from the NASA Astrophysics Theory Program. CV, TJJ, and AKH acknowledge support from the \textit{Fermi} Guest Investigator Program as well as fruitful discussions with Patrick Weltevrede.


%

\end{document}